\documentclass[submission,copyright,creativecommons]{eptcs}
\usepackage{etex}
\usepackage{pstricks, pst-node, pst-tree}
\usepackage{gastex}
\usepackage{stmaryrd}
\usepackage{framed}
\usepackage{amsmath}
\usepackage{amssymb}
\usepackage{wrapfig,lipsum,booktabs}
\usepackage{tikz}
\usepackage{times}
\usetikzlibrary{shapes.geometric} 
\usetikzlibrary{arrows}
\usetikzlibrary{calc}

\newcommand\dist{\mathsf{dist}}

\newcommand\rps{\mathsf{rps}}

\definecolor{pi}{HTML}{EEc6AC} 
\definecolor{ap}{RGB}{176,23,31}

\providecommand{\event}{GandALF 2015} 

\title{Making the Best of Limited Memory in Multi-Player Discounted Sum Games\thanks{%
The project was supported by the EPSRC through grant EP/M027287/1 (Energy Efficient Control).}}
\author{Anshul Gupta
	 \institute{University of Liverpool\\ Liverpool, UK}
		\and
     Sven Schewe  
     \institute{University of Liverpool\\ Liverpool, UK}
		\and
	 Dominik Wojtczak
	 \institute{University of Liverpool\\ Liverpool, UK}
}
\def\titlerunning{Making the Best of Limited Memory in Multi-Player Discounted Sum Games}
\def\authorrunning{A.\ Gupta, S.\ Schewe, and D.\ Wojtczak}

\newenvironment{pproof}{
	{\sc Proof. }}{}
\newtheorem{theorem}{Theorem}

\newtheorem{lemma}[theorem]{Lemma}

\newtheorem{corollary}[theorem]{Corollary}
\newtheorem{definition}[theorem]{Definition}

\begin{document}
	\maketitle

\begin{abstract}
In this paper, we establish the existence of optimal bounded memory strategy profiles in multi-player discounted sum games. We introduce a non-deterministic approach to compute optimal strategy profiles with bounded memory. Our approach can be used to obtain optimal rewards in a setting where a powerful player selects the strategies of all players for Nash and leader equilibria, where in leader equilibria the Nash condition is waived for the strategy of this powerful player.
The resulting strategy profiles are optimal for this player among all strategy profiles that respect the given memory bound, and the related decision problem is NP-complete. We also provide simple examples, which show that having more memory will improve the optimal strategy profile, and that sufficient memory to obtain optimal strategy profiles cannot be inferred from the structure of the game.
\end{abstract}

\section{Introduction}

    \begin{figure*}
    	\centering 
    \begin{tikzpicture}[->,>=stealth',shorten >=1pt]
	\tikzstyle{vertex}=[circle,fill=black!10,minimum size=17pt,inner sep=0pt,font=\sffamily\small\bfseries]
	\tikzstyle{vertex1}=[circle,minimum size=12pt,inner sep=0pt,font=\sffamily\small\bfseries]
	\tikzstyle{vertex2}=[circle,minimum size=12pt,inner sep=0pt,font=\sffamily\small\bfseries]
	\node (4) at (0.2,1) {};
	\node (15) at (0.5,1.3) {$\frac{1}{3}$} ; 
	\node (17) at (-1,-0.22) {$\frac{1}{3}$} ; 
	\node (19) at (3,-0.22) {$\frac{1}{3}$} ;
	\node (1) at (1,1) [vertex,draw]{$1$} ; 
	\node (16) at (-1.3,-0.5) {};
	\node (6) at (1,2) [vertex1,draw]{} ;
	\node (2) at (2.5,-0.5) [vertex,draw] {$2$};
	\node (18) at (3.3,-0.5) {};
	\node (7) at (3,-1.5) [vertex1,draw]{} ;
	\node (5) at (-0.5,-0.5) [vertex,draw] {$3$};		
	\node (8) at (-1,-1.5) [vertex1,draw] {};
	\node (9) at (-0.1,0.5) [vertex1] {$\overrightarrow{0}$};
	\node (10) at (2.1,0.5) [vertex1] {$\overrightarrow{0}$};
	\node (11) at (1.1,-0.8) [vertex1] {$\overrightarrow{0}$};
	\node (12) at (1.6,1.5) [vertex1] {$(2,1,9)$};
	\node (13) at (3.35,-0.9) [vertex1] {$(9,2,1)$};
	\node (14) at (-1.3,-0.9) [vertex1] {$(1,9,2)$};
	
	\path[every node/.style={font=\sffamily\small}]
	(4) edge [right] node[] {} (1)
	(16) edge [right] node[] {} (5)
	(18) edge [left] node[] {} (2)
	(1) edge [right] node[above] {} (2)      
	edge [above] node[] {} (6) 	    
	(2) edge [right] node[] {} (5)
	edge [left] node[] {} (7)	
	(5) edge [left] node[] {} (8)
	edge [above] node[] {} (1)
	(6) edge [loop above] node[] {$\overrightarrow{0}$} (6)
	(7) edge [loop right] node[] {$\overrightarrow{0}$} (7)
	(8) edge [loop left] node[]  {$\overrightarrow{0}$} (8);
	
	\end{tikzpicture}
	\caption{\small \label{fig:memoryless}
		A game with three players and no memoryless Nash (nor leader) equilibrium for discount factor $\lambda = \frac{1}{2}$. The start vertex is picked uniformly at random out of the vertices 1, 2, 3 controlled by players 1, 2, 3, respectively. Each edge is labelled with a reward vector $(r_1,r_2,r_3)$ where $r_i$ is the reward player $i$ gets for traversing that edge.}
\end{figure*}
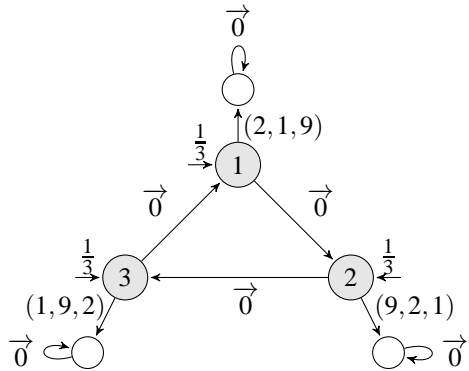

Discounted sum games \cite{shapley:stochastic_games,DBLP:journals/mmor/Sennott94a} are the stochastic games with quantitative objectives that have been introduced by Shapley \cite{shapley:stochastic_games}.
They are played on a finite directed graph without sinks, where each vertex is owned by one of the players.
Intuitively, they are played by placing a token on the graph, which is moved forward by the players.
We consider an initial probability distribution over all vertices to select the start vertex.
As an example, refer to Figure \ref{fig:memoryless}, where vertex $1$, vertex $2$ or vertex $3$ each can be taken as a start (or: initial) vertex with probability $\frac{1}{3}$.
Initially, the token is placed on a start vertex.
Whenever the token is on a vertex, the player who owns this vertex will select an outgoing edge and move the token along this edge.
This way, the players construct an infinite play. Quantitative games \cite{BPS/13/simpleEquilibria} are good models for studying non-terminating programs with multiple components that interact in non-cooperation mode.
In quantitative games, players have goals defined by the payoffs on the edges (sometimes on the vertices). 
For these payoffs, the players have quantitative targets, such as maximising their individual limit average or the discounted sum of their individual rewards, where the value of a play is computed under a discount factor.
Solutions to these games are the strategy profiles that consists of strategies---recipes how to play---for each player.
However, in a realistic situation, these solutions need to be implementable, and thus players have to cope with limited resources such as limited memory. 
Strategy profiles should also satisfy same basic consistency constraints. The reason for this is that the players are assumed to be rational. The lowest level of rationality for a player is to take a look at her strategy profile, and to check if she would gain by changing her own strategy. Strategy profiles where all strategies pass this test are \emph{stable} in terms of Nash equilibria \cite{edselc.2-52.0-000069611119900601,Nash01011950,Osborne1994}. Thus, in a Nash equilibrium, no player benefits from changing her strategy unilaterally. 

The second eminent class of equilibria goes back to von Stackelberg and is referred to as Stackelberg equilibria or \emph{leader equilibria} \cite{von1934marktform}.
In economic game theory, leader equilibria refer to a setting, where a powerful player can move first, or announce her move first, rather than moving at the same time as the remaining players.
This `right of the first move' provides her with some advantage over the other players.
Broadly speaking, the Nash requirements of having no incentive to deviate only affects the remaining players, but not the leader herself.
Leader equilibria have recently been studied as a more general and broader class of strategy profiles than Nash equilibria, called \emph{leader strategy profiles} \cite{6940380}, in the context of multi-player mean payoff games \cite{Zwick+Paterson/96/payoff,DBLP:conf/lics/ChatterjeeHJ05}.
The \emph{leader} can assign the strategies to all players, including herself. While we still require the strategy profile to be stable in that the \emph{other} players do not have an incentive to deviate, the leader herself may be in a position to improve over her current strategy by deviating unilaterally. Thus, every Nash equilibrium is a leader strategy profile, but not every leader strategy profile is Nash. 
We call strategy profiles that are optimal for the leader \emph{leader equilibria (LE)}. The more relaxed condition of a leader strategy profile implies that leader equilibrium can be selected from a 
larger base (cf. Figure \ref{fig:strategyprofiles}). 
The leader's payoff can therefore improve as compared to Nash equilibria.
In this paper, we study leader equilibria and Nash equilibria for the leader in discounted sum games (DSGs) that use bounded memory.

\subsection{Related Work} 
The theory of stochastic games was introduced by Shapley in \cite{shapley:stochastic_games}.
He showed that every two player discounted zero-sum game has a value and that optimal positional strategies exist for both the players.
This idea is further extended in \cite{2053620} to establish the existence of stationary equilibria in stochastic multi-player games.
Bewley and Kohlberg \cite{bewley1978stochastic} have shown that, in two player zero-sum undiscounted stochastic games where both the set of action and the state spaces are finite, stationary optimal strategies exist for both the players.
Gimbert and Zielonka \cite{DBLP:conf/mfcs/GimbertZ04} have studied infinite two player antagonistic games with more general reward functions. They have given sufficient conditions that ensure both the players to have positional (memoryless) optimal strategies. 
Letchford et~al. \cite{DBLP:journals/sigecom/LetchfordMCPI12} have considered computing optimal Stackelberg strategies in stochastic games. They studied this in context with correlation equilibria and discuss the value of correlation and commitment in stochastic games.
Berg and Kitti \cite{berg2013computing} have studied subgame perfect pure strategy equilibria in DSGs. They analyse subgame perfect equilibria in games with perfect information.
Brihaye et~al. \cite{BPS/13/simpleEquilibria} have studied the existence of simple Nash equilibria in non-terminating games with various mixed reward functions.
The strategies used in this paper are inspired by the strategies introduced in \cite{Friedman1971}. 
Gupta and Schewe \cite{6940380} have studied the optimal leader strategy profiles in context with multi-player mean payoff games.

\subsection{Results}
This paper extends the use of leader equilibria to multi-player DSGs.
While all of the above results refer to equilibria that use either none or very small memory---memorising the player who deviated---we show that such simple strategies do not suffice in the case of leader equilibria.
This is owed to mixing the optimality requirements from leader equilibria with discounting. In DSGs, we show that, as a result, the leader can benefit from more memory (Lemma \ref{lem:finitememory}), and that there are actually cases, where infinite memory is needed for leader equilibria 
(Theorem \ref{theo:memorybound} and Theorem \ref{theo:memorybound2}) 
and Nash equilibria (Theorem \ref{theo:memorybound-NE}).

We do not hold strategies that require infinite memory to be realistic, and therefore discuss the construction of strategies that use only bounded memory.
We first show that memoryless leader equilibria do not always exists, a simple corollary from the existence of games without memoryless subgame perfect equilibrium \cite{Kuipers2009}.
The example from Figure \ref{fig:memoryless}, inspired by \cite{Kuipers2009}, has no memoryless Nash equilibria. 
Therefore,  when the leader is not among the three players who own the three central vertices, there is no memoryless leader equilibrium for this game.
There even exists a game with a fixed starting position where no pure Nash equilibria exist \cite{Gurvich2014131}.

This problem, however, seems artificial when reviewing traditional classes of Nash equilibria.
They often use the traditional form of `reward and punish' strategy profiles \cite{Friedman1971,BPS/13/simpleEquilibria,6940380}.
Strategy profiles define a play, the play that ensues when all players follow the strategies assigned to them.
Reward and punish strategy profiles broadly consist of this play, and an agreement that the first player who deviates is punished: all other players collude henceforth, following the new goal to harm the deviator.

Upon deviation, reward and punish strategy profiles therefore turn into two player games, and thus enjoy the usual memoryless determinacy.
The memory needed for this is tiny: one only needs to store who has deviated.
We therefore argue that the resource bounds should refer to the construction of the main play, i.e., main path before deviation.

We give a simple non-deterministic polynomial time approach for assigning reward and punish strategies that meet or exceed a given payoff bound for the leader and uses memory only within a given bound. In Section \ref{sec:constraints}, we show that the decision problem whether a pure strategy with bounded memory that gives a reward greater than or equal to some threshold value exists is NP-complete.

\section{Preliminaries}
A multi-player discounted sum game (MDSG) is a game played on the finite directed weighted graph $\mathcal G$ defined as a tuple 
$\langle P,V,\{V_p\mid p \in P\},\Delta,A,T, \{t_p: V \times A \rightarrow \mathbb Q \mid p \in P\}\rangle$, where $P$ is a finite set of players,
$V$ is a finite set of vertices, 
$\Delta : V \to [0,1]$ is a probability distribution over $V$, which for each $v \in V$ specifies the probability of selecting $v$ as the start vertex.
$\{V_p\mid p \in P\}$ is a partition of the vertices $V$ into the sets $V_p$ of vertices owned by player $p$,
$A$ is a finite set of actions,
$T : V \times A \rightarrow V$ is a set of transitions that maps vertices and actions to vertices,
and $\{t_p \mid p \in P\}$ is a family of reward functions defined as $t_p: V \times A \rightarrow \mathbb Q$ for all $p \in P$ that assigns, for each respective player $p$, a reward for each action~$a$ that is taken from a vertex $v$ (or, likewise, for the transition taken). The game is played by moving a token along the edges of the graph, starting from the start vertex as given by the probability distribution~$\Delta$. We use this initial probability distribution to select a start vertex.
Each vertex $v$ belongs to exactly one player~$p$.
At vertex $v$, the player who owns $v$ selects the next action $a$.
The token is then moved forward to the vertex as given by the transition $T(v,a)$. 
This results in an infinite path, called a \emph{play}.
We denote the reward for player $p$ at any transition $T(v,a)$ by $t_p(v,a)$. An MDSG is called a zero-sum game if, for all vertices $v \in V$ and for all actions $a \in A$, 
$\sum_{p \in P} t_p(v,a) = 0$ holds.
The payoff at every transition is discounted by a discount factor $\lambda$, where $0 < \lambda < 1$.
In DSGs, the payoff (or: reward) for player $p$ at the $i^{th}$ transition is given by $t_p(v_i,a_i)\cdot\lambda^{i}$.
For an infinite play $\pi = v_{0},a_0,v_{1},\ldots$, we denote the reward for player $p$ by 
$r_{p}(\pi) = \sum_{i=0}^{ \infty } t_p(v_i,a_i)\cdot\lambda^{i}$. 

The way that the respective player $p$ chooses the successor vertex is defined by a \emph{strategy} $\sigma_p$. We consider \emph{pure strategies}, which are functions $\sigma_p: (VA)^*V_p \rightarrow A$ from initial sequences of plays to actions.
We focus on two types of pure strategies, memoryless and bounded memory strategies.
A pure memoryless strategy (or: a positional strategy) is a strategy, in which the choice of the next vertex depends only on the current position, whereas a pure bounded memory strategy is a strategy, where the choice of next vertex depends on finite memory.
For a bounded memory $M$ (where $M$ is simply a finite set of fixed size, the memory bound with a dedicated initial value $m_0$), we define two functions: the memory update function, and the memory usage function that provides us with the action that is to be selected.
The memory usage function is a mapping $\mathcal U: M \times V \rightarrow A$ that maps a memory state and a vertex to an action.
In the classic memory model, the memory update function $\mathcal M: M \times (V\times A) \rightarrow M$ defines how the memory is updated; it maps a memory state, a vertex, and an action to a new memory state.
Thus, the memory works as a Moore machine without output, where $M$ is the memory and $\mathcal M$ is the transition function.

As discussed in the introduction, the example from Figure \ref{fig:memoryless} shows that this memory model does not always lead to an equilibrium, at least not for arbitrary $M$.
We therefore define a memory model for reasoning with bounded resources (cf. Corollary \ref{cor:compliance}).
We refer to this model as \emph{compliance memory}, as it only refers to the histories, where all players have complied to their strategies.
This justifies a \emph{partial} memory update function $\mathcal M:  M \times (V\times A) \rightarrow M$, where $\mathcal M(m;v,a)$ is defined if, and only if, $a= \mathcal U(m,v)$.
When the action $a$ differs from the action defined by the memory usage function, the system remembers only who caused the deviation, and then switches into a different mode, where it uses a memoryless strategy (cf. Theorem \ref{theo:optimalPunishment} and Corollary \ref{cor:optimalPunishment}).

The input alphabet $V \times A$ is a product of the last vertex, the action selected, and the vertex reached on a transition. A family of strategies $\sigma = \{\sigma_p \mid p \in P \}$ is called a strategy profile. A strategy profile $\sigma$ defines an expected reward, denoted $\mathbb E_{p}(\sigma)$ for each player $p$. In this paper, we shall focus on the reward of positional and bounded memory strategy profiles. For a positional strategy profile $\sigma$, the payoff from every vertex is well defined. By abuse of notation, we use
$\mathbb E_{p}(\sigma,v) = t_p\big(v,\sigma(v)\big)  + \lambda \mathbb E_{p}\Big(\sigma,T\big(v,\sigma(v)\big)\Big)$ to denote the payoff for player $p$ when starting in a vertex $v$. Note that this implies $\mathbb E_{p}(\sigma)=\sum_{v\in V} \Delta(v) \mathbb E_{p}(\sigma,v)$.

\begin{definition}
[Nash equilibrium]
A strategy profile is a Nash equilibrium if no player has an incentive to change her strategy, provided that all other player keep theirs.
That is, for all players $p \in P$ and for all $\sigma' = (\sigma'_q)_{q \in P}$ with $\sigma_q = \sigma_q'$ for all $q \neq p$, $\mathbb E_p(\sigma) \geq \mathbb E_p(\sigma')$ holds.
\end{definition}

\begin{definition}
[leader strategy profile] 
A strategy profile is a \emph{leader strategy profile} \cite{6940380} for a designated player $l$ (for leader), if no \emph{other} player has an incentive to deviate her strategy.
That is, if, for all players $p \in P\smallsetminus\{l\}$ and for all $\sigma' = \{\sigma_q'\mid q \in P\}$ with $\sigma_q = \sigma_q'$ for all $q \neq p$, $\mathbb E_p(\sigma) \geq \mathbb E_p(\sigma')$ holds.
\end{definition}
A Nash resp.\ leader strategy profile is \emph{optimal} for a class of strategies, if no other strategy profile of this class gives a higher payoff for the leader.

\begin{definition}
[leader equilibrium]
An optimal leader strategy profile for a class of strategies is called a \emph{leader equilibrium}.
\end{definition} 
In two-player DSGs, the set of vertices in $\mathcal G$ is partitioned into two sets where each vertex belongs to exactly one of the players and the player who owns the vertex decides the next move. 
For a MDSG $\mathcal G = \langle P,V,\{V_p\mid p \in P\},\Delta,A,T, \{t_p: V \times A \rightarrow \mathbb Q \mid p \in P\}\rangle$, we define the two-player zero-sum DSG $\mathcal G = \langle P,V,\{V_p,V_o\},\Delta,A,T, \{t_p,t_o\}\rangle$ played between player $p$ and an opponent $o$, where the nodes of $p$ and $o$ partition $V$ into two sets ($V_o =V \smallsetminus V_p$) and their goals are antagonistic $(t_o(v,a) \mapsto -t_p(v,a))$. Note that not all MDSGs with two players in game are two-player games in this sense (two-player games need to be antagonistic zero-sum games). We denote the expected outcome for player $p$ in a two-player game that starts at any vertex $v$ by $r_p(v)$.
A game is called memoryless determined if all players have optimal memoryless strategies. Two-player DSGs are memoryless determined \cite{Zwick+Paterson/96/payoff}: both players have an optimal positional strategy.

\begin{theorem}\cite{Zwick+Paterson/96/payoff}
Two-player DSGs are memoryless determined.
\end{theorem}

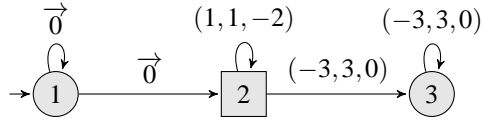
\begin{figure*}
   	\centering 
	\begin{tikzpicture}[->,>=stealth',shorten >=1pt]
	\tikzstyle{vertex}=[circle,fill=black!10,minimum size=17pt,inner sep=0pt,font=\sffamily\small\bfseries]
	\tikzstyle{vertex1}=[rectangle,fill=black!10,minimum size=17pt,inner sep=0pt,font=\sffamily\small\bfseries]
	\node (4) at (0.25,1) {};	
	\node (1) at (1,1) [vertex,draw]{$1$} ; 
	\node (2) at (3.5,1) [vertex1,draw] {$2$};
	\node (3) at (6,1) [vertex,draw] {$3$};
	
	\path[every node/.style={font=\sffamily\small}]
	(4) edge [right] node[] {} (1)
	(1) edge [right] node[above] {$\overrightarrow{0}$} (2)       
	edge [loop above] node[] {$\overrightarrow{0}$} (1)
	(2) edge [right] node[above] {$(-3,3,0)$} (3)
	edge [loop above] node[] {$(1,1,-2)$} (2)
	(3) edge [loop above] node[] {$(-3,3,0)$} (3);     
	
	\end{tikzpicture}
	\caption{\small \label{fig:anshulgame}
	discounted sum game with discount factor $\frac{1}{2}$}
\end{figure*}

\section{Leader and Nash equilibria}
\label{sec:lne}

\begin{figure*}[b]
	\centering
	\begin{tikzpicture}	
	\node [ellipse, ap,thick, minimum height=2.5cm,minimum width=5.5cm, draw] at (-1,0) {};	
	\node [ellipse, red,thick, minimum height=1.5cm,minimum width=3.5cm,draw] at (0,0) {}; 
	\node [ellipse, blue,thick, minimum height=1cm,minimum width=2.5cm,draw] at (0.5,0) {\tiny Nash SPs};
	\node [ellipse, blue,thick, minimum height=1cm,minimum width=2.5cm] at (-1,0.2) {\tiny LSPs};
	\node [ellipse, blue,thick, minimum height=1cm,minimum width=2.5cm] at (-2.7,0.2) {\small General SPs};
	\end{tikzpicture}\vspace*{-2mm}
	\caption{\small \label{fig:strategyprofiles} 
		General strategy profiles $\supseteq$ Leader strategy profiles $\supseteq$ Nash strategy profiles }
\end{figure*}
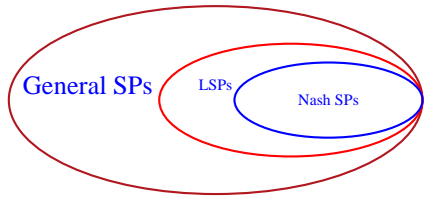

In this section, we show that leader equilibria are superior to Nash equilibria in simple zero-sum DSGs.
For this, consider the three-player game from Figure \ref{fig:anshulgame}. One of the players, player 2, acts as the \emph{leader}. 
The game is played on a simple graph with three vertices, named $1$, $2$, and $3$, owned by the respective player with the same name. 
Note that we denote the vertices owned by leader (resp.\ other players) by square (resp.\ circle) vertices. We used the same notation throughout the paper. In all remaining examples, we select an initial vertex with probability $1$, and therefore mark the initial vertex with an incoming arrow.
The game graph with the payoff vectors of each transition is shown in Figure \ref{fig:anshulgame}, and we use a discount factor of $\lambda = \frac{1}{2}$.
The payoff vectors represent the payoff of player 1, the leader, and player 3, in this order. Initially, player 1 can choose to play to vertex 2 or she can choose to remain in vertex 1. She plays to vertex 2 only if the leader, in her strategy profile, chooses to remain in vertex 2 for a while. At vertex $2$, the leader has different options.

She can choose to play to vertex $3$ (this is the option where she maximises her reward), she can choose to remain in $2$ for a while, before continuing to vertex $3$, or she can stay in vertex $2$ forever. It is easy to notice that, when in vertex $2$, the leader will immediately continue to vertex 3 in all Nash equilibria. Consequently, player $1$ would never play to vertex $2$ from vertex $1$: staying in vertex $1$ forever will yield a payoff of $0$, while moving to vertex $2$ in round $i$ would, for $\lambda = \frac{1}{2}$, result in a payoff of $-\frac{3}{2^i}$. Thus, the only play that can result from a Nash equilibrium is the play $1^\omega$, where the overall reward for all participating players is $0$. However, in a leader equilibrium the leader stays twice in vertex $2$ and then progresses to vertex $3$.
In this case, the leader can assign player $1$ the strategy to immediately progress to vertex $2$, resulting in the play $1,2,2,2,3^\omega$. This will provide an overall payoff of
$0$ for player $1$, $1.5$ for the leader, and $-1.5$ for player~$3$.

\begin{figure*}[thbp]
	\centering
	\begin{tikzpicture}[->,>=stealth',shorten >=1pt]
	\tikzstyle{vertex}=[circle,fill=black!10,minimum size=17pt,inner sep=0pt,font=\sffamily\small\bfseries]
	\tikzstyle{vertex1}=[rectangle,fill=black!10,minimum size=17pt,inner sep=0pt,font=\sffamily\small\bfseries]
	\node (4) at (1,1.75) {};	
	\node (1) at (1,1) [vertex,draw]{$1$}; 
	\node (2) at (2.5,1) [vertex1,draw] {$2$};
	\node (3) at (4,1) [vertex,draw] {$3$};
	\node (5) at (-0.5,1) [vertex,draw] {$4$};
	
	\path[every node/.style={font=\sffamily\small}]
	(4) edge [right] node[] {} (1)
	(1) edge [right] node[above] {$\overrightarrow{0}$} (2)       
	edge [below] node[above] {$\overrightarrow{0}$} (5)
	(2) edge [right] node[above] {$\overrightarrow{0}$} (3)
	edge [loop above] node[] {1,0} (2)
	(3) edge [loop above] node[] {0,50} (3)
	(5) edge [loop above] node[] {$1-\epsilon,0$} (5);
	
	\end{tikzpicture}
	\caption{\label{fig:finitegain}
		increasing the memory helps}
\end{figure*}
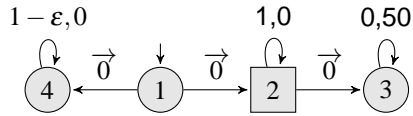

\begin{theorem}
\label{theo:leaderequilibria}
Compared to Nash equilibria, leader equilibria may result in higher, but will never provide smaller rewards for the leader. 
\end{theorem}

\noindent\emph{Proof. }
While the example has proven the `higher' part, note for the `not smaller' part that all Nash equilibria are leader equilibria, such that a leader equilibrium cannot be inferior to a Nash equilibrium. They can, of course, be equal when a leader equilibrium is Nash. 
This is, for example, the case when leader owns no vertex. Thus, leader equilibria gives more leeway to the leader for the selection of optimal strategy profiles and forms a larger base of strategy profiles to choose from, as shown in Figure \ref{fig:strategyprofiles}.

\medskip
Note that the game from Figure \ref{fig:anshulgame} can be used to argue that having memory helps, and having more memory helps more. Among the positional strategies of the leader, staying in vertex 2 forever (with an overall payoff of $1$ for player $1$ and the leader, and $-2$ for player 3, respectively) is superior to continuing immediately to vertex 3 (because in the latter case player $1$ will stay in vertex $1$, see above). So, while still superior to the only Nash equilibrium, it is inferior to the strategy described above, which uses a tiny amount of memory. To observe that, in general, more memory helps more, consider the situation where one lets $\lambda$ grow towards one. It is easy to see that, the closer $\lambda$ gets to one, the longer leader would stay in vertex $2$ in leader equilibrium for the respective discount factor. The optimal memory bounded strategy for the leader therefore improves with the memory we allow for.

\begin{lemma}
\label{lem:finitememory}
The optimal reward for the leader in a Nash or leader equilibrium improves with the increase of the available memory.
\end{lemma}
It now becomes tempting to assume that we could use this observation to identify a situation where an optimal leader strategy profile is reached. That is,
given a fixed discount factor, is there a $k \in \mathbb N$ such that an optimal leader strategy profile for memory $k$ is considered optimal for infinite memory?
The answer to this question is negative. 

\begin{theorem}
\label{theo:memorybound}
For any fixed discount factor $\lambda$, there is no memory bound $k$ such that an optimal leader strategy profile with memory bound $k$ is an optimal leader strategy profile. 
\end{theorem}
\noindent\emph{Proof. }
For this, we refer to the example from Figure \ref{fig:finitegain}, where leader acts as player~2. 
Here, we argue that having a finite memory at the vertices is sufficient for a leader equilibrium, but the effect of increasing the memory is different than in our first example. Irrespective of the discount factor it is apparent that the leader needs to promise sufficiently many, say $s$, loops in vertex $2$ so that $\sum_{i=0}^{s-1} \lambda^i \geq \frac{1-\varepsilon}{1-\lambda}$. Consequently, the number of repetitions grows to infinity, for all $\lambda \in ]0,1[$, and with $\varepsilon$ falling to $0$. 
If the memory is smaller than minimal such $s$, then the leader would receive an overall reward of $0$, either because she promises to stay for more than the memory bound many steps (and thus for ever) in vertex $2$, or by not promising to do so and hence tempting the first player to move to vertex~$4$. If, on the other hand, the memory size is at least $s$, then the leader has enough memory to play the optimal pure strategy to move to vertex $3$ after $s$ loops in vertex $2$.

\medskip
Finally, so far for a fixed discount factor and a fixed game graph with weights, bounded memory was sufficient to guarantee optimal reward to the leader. We now show that infinite memory is sometimes needed in a leader equilibrium.

\begin{figure*}[hbtp]
	\centering
	\begin{tikzpicture}[->,>=stealth',shorten >=1pt]
	\tikzstyle{vertex}=[circle,fill=black!10,minimum size=17pt,inner sep=0pt,font=\sffamily\small\bfseries]
	\tikzstyle{vertex1}=[rectangle,fill=black!10,minimum size=17pt,inner sep=0pt,font=\sffamily\small\bfseries]
	\node (4) at (1,1.75) {};	
	\node (1) at (1,1) [vertex,draw]{$1$} ; 
	\node (2) at (2.5,1) [vertex1,draw] {$2$};
	\node (5) at (-0.5,1) [vertex,draw] {$3$};		
	\path[every node/.style={font=\sffamily\small}]
	(4) edge [right] node[] {} (1)
	(1) edge [right] node[above] {$1,0$} (2)       
	edge [below] node[above] {$\overrightarrow{0}$} (5)
	(2) edge [loop right] node[] {$-1,1$} (2)
	edge [loop above] node[] {$\overrightarrow{0}$} (2)
	(5) edge [loop above] node[] {$\overrightarrow{0}$} (5);
	\end{tikzpicture}
	\caption{\small \label{fig:infinite}
		leader benefits from infinite memory}
\end{figure*}
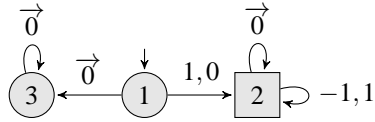

\begin{theorem}
\label{theo:memorybound2}
Optimal leader strategy profile may require infinite amount of memory even for a fixed two-player game
with a fixed discount factor.
\end{theorem}
\noindent\emph{Proof. }
We show this for a two-player game with  
three vertices depicted in Figure \ref{fig:infinite}, where the leader is player~2.
Vertices~1 and 3 belong to player~1 and vertex~2 belongs to the leader. 
The rewards are depicted in the order (player 1, player 2) and we set $\lambda = 2/3$. Notice that player 1 will move to vertex 3 unless the leader can guarantee him a reward $\geq -1/\lambda = -3/2$ from vertex 2, 
because only then his total reward would be $\geq 1-\lambda/\lambda = 0$. On the other hand, in the optimal leader strategy, the leader will try to give him exactly that much, because only then her payoff would be equal to $\lambda/\lambda = 1$. Proposition 1 in \cite{DBLP:conf/lpar/ChatterjeeFW13} shows that the leader can achieve this value with a pure strategy, but only if she has an infinite amount of memory. 

We can show the same for Nash equilibrium, but with 3-players. 
Also, we show that the optimal payoff of a player cannot be
approximated by considering strategies with
bounded memory only.

\begin{figure*}[hbtp]
	\centering
	\begin{tikzpicture}[->,>=stealth',shorten >=1pt]
	\tikzstyle{vertex}=[circle,fill=black!10,minimum size=17pt,inner sep=0pt,font=\sffamily\small\bfseries]
	\tikzstyle{vertex1}=[rectangle,fill=black!10,minimum size=17pt,inner sep=0pt,font=\sffamily\small\bfseries]
	\tikzstyle{vertex2}=[circle,fill=black!10,minimum size=12pt,inner sep=0pt,font=\sffamily\small\bfseries]
	
	\node (6) at (0.25,1) {};
	\node (1) at (1,1) [vertex,draw]{$1$} ;
	\node (2) at (3.5,1) [vertex,draw] {$2$};
	\node (4) at (1,0) [vertex2,draw] {};
	\node (3) at (6,1) [vertex1,draw] {$3$};
	\node (5) at (3.5,0) [vertex2,draw] {};
	\node (7) at (4.75,1.25) {\small $(-1,-1,1)$};
	\node (8) at (2.25,1.25) {\small $\overrightarrow{0}$};
	\node (9) at (1.27,0.5) {\small $\overrightarrow{0}$};
	\node (10)at (3.75,0.5) {\small $\overrightarrow{0}$};
	
	\path[every node/.style={font=\sffamily\small}]
	(6) edge [right] node[] {} (1)
	(1) edge [right] node[] {} (2)      
	edge [below] node[below] {} (4)     
	(2) edge [below] node[below] {} (5)
	edge [right] node[] {} (3)
	(3) edge [loop above] node[] {$(1,0,1)$} (3)
	    edge [loop below] node[] {$(0,1,1)$} (3)       
	(4) edge [loop below] node[] {$\overrightarrow{0}$} (4)
	(5) edge [loop below] node[] {$\overrightarrow{0}$} (5);
	
	\end{tikzpicture}
	\caption{\small \label{fig:Nashinfinite}
		leader benefits from infinite memory in Nash equilibria}
\end{figure*}
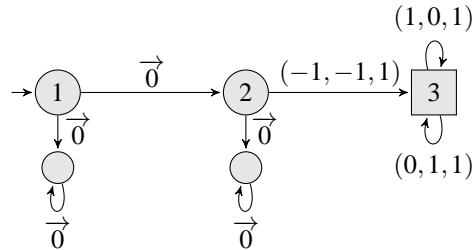

\begin{theorem}
\label{theo:memorybound-NE}
An optimal Nash equilibrium may require infinite amount of memory even for a fixed 3-player game with a fixed discount factor.
Moreover, leader's optimal payoff can be arbitrary far away from her optimal payoff for bounded memory strategies.
\end{theorem}
\noindent\emph{Proof. }
To show this, we refer to the Figure \ref{fig:Nashinfinite}. We have three players here -- player 1, player 2 and leader. 
The vertex~$1$, vertex~$2$ and vertex~$3$ are owned by player~$1$, player~$2$ and leader respectively. 
Rewards are given on the edges and are shown in the order (player 1, player 2, leader). We set the value of discount factor to be $\lambda = 2/3$. Starting from the initial vertex (vertex 1), player 1 can either go to the terminal state that has a reward of 0 for all the three players, or can move to the vertex 2. Similarly, at vertex 2, player 2 can either go to the terminal state or move to the leader vertex. 

For an optimal strategy profile, leader has to promise to both player 1 and player 2 a reward of at least $3/2$ at vertex 3, as otherwise at least one of them would prefer to terminate the game 
at their respective vertices. 
On the other hand, no matter what leader does, their rewards at vertex 3 sum up to $3$,
because the sum of their payoffs on the edges from vertex 3 is constant and equal to $1$.
Therefore, the leader has to promise to both player 1 and player 2 a reward of exactly $3/2$.
Proposition 1 in \cite{DBLP:conf/lpar/ChatterjeeFW13} shows that the leader can achieve this value with a pure strategy, but only if she has an infinite amount of memory. 
The overall rewards of player 1 and player 2 from such a play $1 \cdot 2 \cdot 3^\omega$ would be 0. 
Note that this strategy profile would be a Nash equilibrium where leader's payoff is $2$. 

Finally, if the leader has only bounded amount of memory then one of the other players has to receive less than 
0 from a play $1 \cdot 2 \cdot 3^\omega$ and would prefer to terminate the game before it reaches
vertex 3.
This implies that the optimal payoff of the leader for bounded strategies is $0$, while for 
general strategies it is $2$.
The difference between these two can be made arbitrarily large by scaling the payoffs on 
the edges in this game.

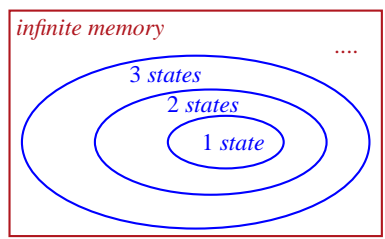
\begin{figure*}[btp]
  	\centering
	\begin{tikzpicture}
	{\footnotesize		
		\node [ellipse, blue,thick, minimum height=.7cm,minimum width=1.54cm, draw] at (0.8,0){};
		\node [ellipse, blue,thick, minimum height=1.4cm,minimum width=3.08cm,draw] at (0.6,0) {};
		\node [ellipse, blue,thick, minimum height=2.3cm,minimum width=4.62cm,draw] at (0.4,0) {};
		\node [rectangle, ap,thick, minimum height=3cm,minimum width=4.95cm,draw] at (0.4,0.25) {};
		
		\node [ellipse, blue,thick, minimum height=1.4cm,minimum width=2.5cm] at (0.9,0){\footnotesize   \emph{$1$ state}};
		\node [ellipse, blue,thick, minimum height=1.4cm,minimum width=2.5cm] at (0.5,0.5){\footnotesize \emph{$2$ states}};
		\node [ellipse, blue,thick, minimum height=1.4cm,minimum width=2.5cm] at (0,0.9){\footnotesize \emph{$3$ states}};
		\node [ellipse, ap,thick, minimum height=1.4cm,minimum width=2.5cm] at (2.4,1.2){....};
		\node [ellipse, ap,thick, minimum height=1.4cm,minimum width=2.5cm] at (-1,1.5){\footnotesize \emph{infinite memory}};;
	}
	\end{tikzpicture}
	\caption{\label{fig:memory}\small more memory states $\Rightarrow$ more strategies}
\end{figure*}

Thus, an optimal strategy profile for a given player can be formed from a memoryless strategy, finite memory strategy or from infinite memory.
More memory would, therefore, give more leeway to leader to select an optimal strategy profile (cf. Figure \ref{fig:memory}).

\section{Reward and punish strategy profiles in discounted sum games}
\label{sec:rps}
In this section, we show that for a play $\pi$, we could establish if there exists a leader (or Nash) strategy profile $\sigma$ with $\pi = \pi_\sigma$, and, moreover, its extension to such a strategy profile is simple. For this, we first introduce reward and punish strategy profiles.

In reward and punish strategy profiles \cite{6940380}, the leader assigns a strategy to each player and each of them co-operates to produce a play $\pi$ while playing in accordance with the assigned strategies.
As soon as one player deviates, the remaining players team up with the leader and co-operate against the deviating player $i$.
That is, they will henceforth follow the goal to minimise the payoff of player $i$, and act jointly as the antagonist of $i$ in the underlying two-player DSG. Thus, in the resultant two-player game, while the objective of player $i$ is still the same, the objective of all other players (including leader), is changed and has become to minimise the payoff of player $i$.
Assuming that positional optimal strategies in this two-player DSG are fixed, $\pi$ thus defines a reward and punish strategy profile, which we denote by $\rps(\pi)$. We now argue that

\begin{enumerate}
\item every leader resp.\ Nash strategy profile $\sigma$ can be transformed into a leader resp.\ Nash strategy profile $\sigma'$ with $\pi_\sigma = \pi_{\sigma'}$, and thus with similar rewards for all players, and
\item give necessary and sufficient conditions for a play $\pi$ to be defined by some leader resp.\ Nash strategy profile.
\end{enumerate}     

We first discuss the necessary conditions for a path to be the outcome of a Nash (resp.\ leader) equilibrium, and then show that it is sufficient for a path to be the outcome of a Nash (resp.\ leader) reward and punish strategy profile.

\begin{lemma}
\label{lem:constraints}
If $\pi = v_0,a_0,v_{1},\ldots$ is the outcome of a Nash (resp.\ leader) equilibrium, then, for all $j \in \mathbb N$ and all players $p$ (resp.\ all players $p \neq l$), $r_{p}(v_j) \leq \sum_{i=0}^{ \infty } t_p(v_{j+i},a_{j+i})\cdot\lambda^{i}$ holds.
\end{lemma}

\begin{pproof}
We assume for contradiction that the condition is violated.
We therefore select a $j \in \mathbb N$, and a player $p$ (for leader equilibria a player $p \neq l$) such that 
$r_{p}(v_j) > \sum_{i=0}^{ \infty } t_p(v_{j+i},a_{j+i})\cdot\lambda^{i}$.
We then change the strategy of player $p$ to follow her strategy from the two player discounted-sum game from position $j$ onwards.
The resulting play $\pi' = v_0',a_0',v_{1}',\ldots$ with $v_i' = v_i$ for all $i \leq j$ and $a_i' = a_i$ for all $i<j$ satisfies
\begin{description}
 \item [$r_{p}(\pi')=$] $\sum_{i=0}^{ \infty } t_p(v_{i}',a_{i}')\cdot\lambda^{i}$
 $=$
  $\sum_{i=0}^{j-1} t_p(v_{i}',a_{i}')\cdot\lambda^{i} \, + \, \lambda^{j}\sum_{i=0}^{ \infty } t_p(v_{j+i}',a_{j+i}')\cdot\lambda^{i}$
 \item [$\geq$] $\sum_{i=0}^{j-1} t_p(v_{i},a_{i})\cdot\lambda^{i} \, + \, \lambda^{j}r_{p}(v_j)$
 $>$
  $\sum_{i=0}^{j-1} t_p(v_{i},a_{i})\cdot\lambda^{i} \, + \, \lambda^{j}\sum_{i=0}^{ \infty } t_p(v_{j+i},a_{j+i})\cdot\lambda^{i}$
 \item [$=$] $\sum_{i=0}^{ \infty } t_p(v_{i},a_{i})\cdot\lambda^{i} = r_{p}(\pi)$. 
\end{description}
\end{pproof}

\begin{lemma}
If $\pi = v_0,a_0,v_{1},\ldots$ satisfies 
$r_{p}(v_j) \leq \sum_{i=0}^{ \infty } t_p(v_{j+i},a_{j+i})\cdot\lambda^{i}$
for all $j \in \mathbb N$ and all players $p$ (resp.\ all players $p \neq l$), then
$\rps(\pi)$ is a Nash (resp.\ leader) equilibrium.
\end{lemma}

\begin{pproof}
We assume for contradiction that a player $p$ (for leader equilibria a player $p \neq l$) has an incentive to deviate, and that the first position where player $p$ selects a different action is $j \in \mathbb N$.
Let $\pi' = v_0,a_0',v_{1}',\ldots$, where $v_i' = v_i$ for all $i \leq j$ and $a_i' = a_i$ for all $i<j$, be the resulting play. We have,

\begin{description}
 \item [$r_{p}(\pi')=$] $\sum_{i=0}^{ \infty } t_p(v_{i}',a_{i}')\cdot\lambda^{i}$
 $=$
  $\sum_{i=0}^{j-1} t_p(v_{i}',a_{i}')\cdot\lambda^{i} \, + \, \lambda^{j}\sum_{i=0}^{ \infty } t_p(v_{j+i}',a_{j+i}')\cdot\lambda^{i}$
 \item [$\leq$] $\sum_{i=0}^{j-1} t_p(v_{i},a_{i})\cdot\lambda^{i} \, + \, \lambda^{j}r_{p}(v_j)$
 $\leq$
 $\sum_{i=0}^{j-1} t_p(v_{i},a_{i})\cdot\lambda^{i} \, + \, \lambda^{j}\sum_{i=0}^{ \infty } t_p(v_{j+i},a_{j+i})\cdot\lambda^{i}$
 \item [$=$] $\sum_{i=0}^{ \infty } t_p(v_{i},a_{i})\cdot\lambda^{i} = r_{p}(\pi)$. 
\end{description}
\end{pproof}

The first `$\leq$' is implied by the definition of $\rps$, as the remaining players will play antagonistic to $p$, such that $p$ cannot yield a better result than $r_{p}(v_j)$ starting from $v_j$.
Together with the observation that pure Nash equilibria always exist \cite{BPS/13/simpleEquilibria}---leader equilibria can be formed by all players (playing as if they played their respective two-player discounted sum game)---these lemmas provide the following theorem.

\begin{theorem}
\label{theo:optimalPunishment}
Pure Nash and leader strategy profiles always exist in MDSGs, and for finding optimal ones, it suffices to consider reward and punish strategies.
\end{theorem}

This is particularly interesting when we focus on the implementable strategy profiles.
A strategy is \emph{implementable}, if it is realisable with finite memory.
We are particularly interested in finite memory strategies with a given small bound $b$ on the memory used.
Note that, for reward and punish strategy profiles, we do not have to record the reaction upon deviation, as it is implicitly described by the punishment part.
Thus, we do not want to reason about the trivial part in the strategy, and therefore do not count the tiny bit of memory required for the punishment part.
This part does not need much memory: it suffices to memorise which player is responsible for the deviation and at which vertex. When we allow for finite memory $M$, this effectively defines a larger game, on which a memoryless strategy is used. For a game $\mathcal G = \langle P,V,\{V_p\mid p \in P\}, \Delta, A, T, \{t_p: V \times A \rightarrow \mathbb Q \mid p \in P\}\rangle$ and finite memory $M$ with initial memory $m_0 \in M$, we can simply define 
$\mathcal G^M = \langle P,V',\{V_p'\mid p \in P\}, \Delta',A,T', \{t_p': V' \times A \rightarrow \mathbb Q \mid p \in P\}\rangle$ with
$V' = V \times M$, $V_p' = V_p \times M$, $T' : V' \times A \rightarrow V'$ is a set of transitions that maps vertices and actions to vertices, $\Delta'(v,m) = \Delta(v)$ if $m = m_0$ and $\Delta'(v,m) = 0$ otherwise, and $t_p' : ((v,m),a) \mapsto t_p(v,a)$.

\begin{corollary}
\label{cor:optimalPunishment}
Pure memoryless, and, for a given memory bound $b$, pure bounded memory Nash and leader strategy profiles always exist in MDSGs, and for finding the optimal ones, it suffices to consider reward and punish strategies.
\end{corollary}

\begin{corollary}
\label{cor:compliance}
For optimal reward and punish strategy profiles, it suffices to consider the memory needed before deviation, i.e., compliance memory and additional $k$ memory states for the $k$ followers, 
rather than considering an arbitrary memory $M$.
\end{corollary}

\section{Constraints for finite pure reward and punish strategy profiles}
\label{sec:constraints}
We first state that optimal strategies exist for all memory bounds.
This is a simple implication of Theorem \ref{theo:optimalPunishment} and the finite space of candidate strategy profiles.

\begin{lemma}
\label{lem:optExist}
For all MDSGs and for all memory bounds, optimal strategy profiles exist among the Nash and leader equilibria.
\end{lemma}

We infer a necessary and sufficient constraint system for the strategy profiles in Nash and leader equilibria in MDSGs. 
Theorem \ref{theo:optimalPunishment} implies that, whenever a player deviates at some vertex $v$, then the remainder of the game resembles a two-player game that starts at $v$. The player who owns vertex $v$ therefore has an incentive to deviate if, and only if, her payoff from now onwards would be less than the payoff she receives in this underlying two-player game. 
This provides us with a first necessary constraint, namely
\begin{itemize}
	\item at any history $h$ that ends in a vertex $v$ 
	 owned by player $p \in P$,
	$\mathbb E_{p}(\sigma,h) \geq  r_p(v)$. 
\end{itemize}

For positional (or: memoryless) reward and punish strategies
$\sigma$, the subtrees in all histories $h$ that end in $v$ coincide, such that one can write $\mathbb E_{p}(\sigma,v)$ instead of 
$\mathbb E_{p}(\sigma,h)$.

For pure strategies, we require 
for every vertex $v$ that
\begin{itemize}
\item for all players $p \in P$,
$\mathbb E_{p}(\sigma,v) = t_p\big(v,\sigma(v)\big)  + \lambda \mathbb E_{p}\Big(\sigma,T\big(v,\sigma(v)\big)\Big)$.
\end{itemize}
The action $\sigma(v)$ from these constraints refers to the action selected at vertex $v$ by player $p$ in strategy profile $\sigma$.
Once these actions are fixed, we therefore have a simple linear equation system of full degree, that can easily be solved.
To determine if the resulting system is in equilibrium we can simply check if the first set of constraints hold for all players (Nash equilibrium) or for all players but the leader (leader equilibrium). To validate that there is a pure strategy profile of a predefined quality can therefore be checked in nondeterministic polynomial time.

\begin{lemma}
\label{lem:inNP}
We can check, if there is a positional strategy profile that meets or exceeds a given threshold $t$ for the leader reward and is a leader or Nash equilibrium, in nondeterministic polynomial time.
\end{lemma}

For strategy profiles with bounded memory, we can simply use the extended memory game instead. We can also prove NP hardness of this problem using standard reduction from 3-SAT as in \cite{UW11,DBLP:conf/fossacs/Ummels08}. By putting these two together we obtain the following theorem.

\begin{theorem}
	\label{theo:NPC}
	To check, if there is a pure positional or bounded memory strategy profile with fixed memory bound $b$ that meets or exceeds a given threshold $t$ for the leader and is a leader or Nash equilibrium, is NP complete.
\end{theorem}

\begin{figure*}
	\centering
	\begin{tikzpicture}[->,>=stealth',shorten >=1pt,auto,node distance=1.2 cm, thick,
	path node/.style ={thin,circle,fill=pi,draw,font=\sffamily\tiny},main node/.style={thin,circle,fill=black!6,draw,font=\sffamily\tiny}, second node/.style={circle,font=\sffamily\tiny}, third node/.style={circle,node distance = 0.6 cm, font=\sffamily\tiny}, fourth node/.style={circle,node distance = 0.7 cm, font=\sffamily\tiny},conjunct node/.style={thin,rectangle,rounded corners,
		minimum height = 3cm,minimum width = 1cm,draw,font=\sffamily\tiny}, conjunct node1/.style={thin,rectangle,rounded corners, minimum height = 3.2cm,minimum width = 1cm,draw,font=\sffamily\tiny},main node1/.style={thin,rectangle,fill=black!6,draw,font=\sffamily\tiny},
	conjunct node2/.style={thin,rectangle,rounded corners, minimum height = 3.1cm,minimum width = 1cm,draw,font=\sffamily\tiny},conjunct node3/.style={thin,circle,font=\sffamily\large},
	dotted node/.style={thin,circle,fill=black,draw,font=\sffamily\tiny}]
	
	\node[path node] (1) {$p$};
	\node[conjunct node] (16) at (0,-1.22) {$c_1$};
	\node[conjunct node1] (17) at (2.4,-1.2) {$c_2$};
	\node[conjunct node2] (18) at (4.8,-1.2) {$c_3$};
	\node[main node] (2) [below of=1] {$\neg q$};
	\node[main node] (3) [below of=2] {$\neg r$};  
	\node[conjunct node3] (19) at (-0.9,-2.5) {$C_1$};
	\node[conjunct node3] (20) at (1.5,-2.5) {$C_2$};
	\node[conjunct node3] (20) at (3.9,-2.5) {$C_m$};
	\node[main node1] (4) [above right of=3,node distance=1.7cm] {$L$};  
	\node[main node] (5) [above right of=4,node distance=1.7cm] {$\neg p$};
	\node[path node] (6) [below of=5] {$q$};
	\node[main node] (7) [below of=6] {$\neg r$};  
	\node[main node1] (8) [above right of=7,node distance=1.7cm] {$L$};
	\node[main node] (9) [above right of=8,node distance=1.7cm] {$\neg p$};
	\node[main node] (10) [below of=9] {$\neg q$};
	\node[path node] (11) [below of=10] {$\neg r$};
	\node[main node1] (12) [below left of=1,node distance=1.7cm] {$L_s$};
	\node[main node1] (13) [above right of=11,node distance=1.7cm] {$L_s$};
	\node[main node] (14)[above of=5,node distance=1cm]{$abs$};
	\node[main node] (15)[below of=7,node distance=1cm] {$abs$};
	\node (16) at (-2,-1.22) {};
		
	\path[every node/.style={font=\sffamily\small}]
	(1)  edge [pi, thin] node [below right] {} (4) 
	     edge [thin, dashed] node [right] {} (14)
	(2)  edge [thin] node [] {} (4)		
	     edge [thin, dashed] node [left] {} (14)
	(3)  edge [thin] node [above right] {} (4)
	     edge [thin, dashed] node [right] {} (15)    
	(4)  edge [thin] node [above right] {} (5)
   	     edge [pi,thin] node [] {} (6)
	     edge [thin] node [below right] {} (7)    
	(5)  edge [thin] node [below right] {} (8)    
	     edge [thin, dashed] node [above] {} (14)
	(6)  edge [pi,thin] node [] {} (8)		 
	     edge [thin, dashed, bend left] node [below] {} (15)
	(7)  edge [thin] node [above right] {} (8)        
	     edge [thin, dashed] node [below] {} (15)
	(8)  edge [thin] node [above right] {} (9)
	     edge [thin] node [] {} (10)
	     edge [pi,thin] node [below right] {} (11)    
	(9)  edge [thin, dashed] node [left] {} (14)
	     edge [thin] node [below right] {} (13)
	(10) edge [thin] node [right] {} (13)
	     edge [thin, dashed] node [left] {} (14)	
	(11) edge [pi,thin] node [above right] {} (13)
	     edge [thin, dashed] node [left] {} (15)
	(12) edge [pi, thin] node [above right] {} (1)
	     edge [thin] node [] {} (2)
	     edge [thin] node [] {} (3)
	(14) edge [loop left, thin, dashed] node[] {} (14)   
	(15) edge [loop right,thin, dashed] node[] {} (15)  
	(16) edge [thin] node[] {} (12)  ;
	\end{tikzpicture}
	\caption{\small \label{fig:validationphase}
		$C_1, C_2 ...... C_m$ are $'m'$ conjuncts each with $'n'$ variables and there are intermediate leader $'L'$ nodes.
		A path through the satisfying assignment is shown here.}	
\end{figure*}
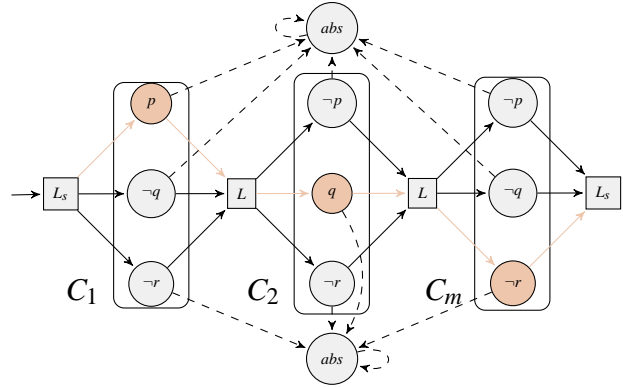

\begin{pproof}	
In order to establish NP completeness, we reduce the satisfiability of a 3SAT formula $\varphi$ over $n$ atomic propositions with $m$ conjuncts to solving a multi-player discounted sum game with $2n+1$ players and $4m+5n+2$ vertices that uses only payoffs $-1$ and $0$. 
Note that we have not considered discount factor in the proof. 
We gave a standard reduction, which is similar to the reduction for mean payoff games \cite{6940380}, safe for the weights.

We consider the reduction for the example of the 3SAT formula $(p \vee \neg q \vee \neg r) \wedge (\neg p \vee q \vee \neg r) \wedge (\neg p \vee \neg q \vee \neg r)$.
The $2n+1$ players consists of $2n$ players for the $2n$ literals corresponding to the $n$ variables, and the \emph{leader}, who intuitively tries to validate the formula.
The vertices are labelled by their owner. 

The payoff for a transition that goes from a vertex owned by a literal player $l$ to a vertex different to the absorbing state `abs' has a payoff of $-1$ for the player $\neg l$, and of $0$ for every other player. The self-loop at `abs' has a payoff of $-1$ for the leader, and of $0$ for every other player.
The remaining transitions have payoffs of $0$ for every player. 

If $\varphi$ is satisfiable, the leader can use a satisfying assignment to determine a cycle through the game graph that does not pass by two vertices owned by opposing literal players $p$ and $\neg p$. 
All players that make a decision in the unfolding infinite path have a reward of $0$, which is the optimal reward obtainable in any play, as there are no positive rewards on any edge. In this case, the leader reward is $0$. 

Let us assume that $\varphi$ is unsatisfiable, and the play defined by the leader in a leader strategy profile does not end in the absorbing state. Then there is a first literal $l$ on the play, whose negation $\neg l$ occurs later. The player who owns $l$ will receive a negative return when complying, and hence deviate by moving to the absorbing state. This way, the player receives a reward of $0$. 
Hence, every play in a leader equilibrium for unsatisfiable assignments must end in the absorbing state, which implies that the leader receives a negative reward.

The example is depicted in the Figure \ref{fig:validationphase}. There are total $m$ conjuncts and each conjunct has $n$ literal variables.
Thus, for $n$ propositions, there are $2n$ literal variables. 
We refer to leader nodes as $'L'$, leader's starting node as `$L_s$' and there is one absorbing state `abs'.
`$L_s$' is taken as start node with probability $1$.
The two depicted copies of the vertices `abs' and `$L_s$' each refer to one vertex. 
As inclusion in NP has been shown in Lemma \ref{lem:inNP} for positional strategies and we can simply use the extended memory game instead, we infer NP completeness.

\end{pproof}
	
\section{Equilibria with extended observations}
\label{sec:strange}

We first argue why we have focus on the pure strategies only in the previous sections, although mixed strategies are a more general choice.
In principle, all arguments from the previous sections also extend to the randomised strategies and strategy profiles, such that one might argue to use the randomised model.
The reason why we refrained from doing so is that reward and punish strategies rely on the observability of deviation.

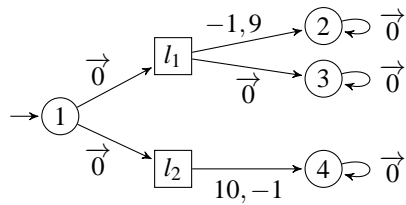
\begin{figure*}[h]
	\centering 
	\begin{tikzpicture}[->,>=stealth',shorten >=1pt]
	\tikzstyle{vertex}=[circle,fill=black!10,minimum size=17pt,inner sep=0pt,font=\sffamily\small\bfseries]
	\tikzstyle{vertex1}=[circle,minimum size=14pt,inner sep=0pt,font=\sffamily\small\bfseries]
	\tikzstyle{vertex2}=[rectangle,minimum size=14pt,inner sep=0pt,font=\sffamily\small\bfseries]
	\node (4) at (0.2,1) {};	
	\node (1) at (1,0.3) [vertex2,draw]{$l_1$} ; 
	\node (16) at (-1.3,-0.5) {};
	\node (6) at (3,0.7) [vertex1,draw]{$2$} ;
	\node (2) at (1,-1.2) [vertex2,draw] {$l_2$};
	\node (18) at (3.3,-0.5) {};
	\node (7) at (3,-1.2) [vertex1,draw]{$4$} ;
	\node (5) at (-0.5,-0.5) [vertex1,draw] {$1$};
	\node (3) at (3,0) [vertex1,draw]{$3$} ;		
	\node (8) at (2,-0.15) [vertex1] {$\overrightarrow{0}$};
	\node (9) at (0,0.1) [vertex1] {$\overrightarrow{0}$};
	\node (10) at (0,-1.1) [vertex1] {$\overrightarrow{0}$};
	\node (11) at (2,-1.5) [vertex1] {$10,-1$};
	\node (12) at (1.8,0.7) [vertex1] {$-1,9$};
		
	\path[every node/.style={font=\sffamily\small}]
	(16) edge [right] node[] {} (5)	
	(1) edge [below] node[] {} (3)      
	(1) edge [above] node[] {} (6) 	    
	(2) edge [left] node[] {} (7)	
	(3) edge [loop right] node[] {$\overrightarrow{0}$} (3)	
	(5) edge [above] node[] {} (1)
	edge [above] node[] {} (2)
	(6) edge [loop right] node[] {$\overrightarrow{0}$} (6)
	(7) edge [loop right] node[] {$\overrightarrow{0}$} (7);
		
	\end{tikzpicture}
	\caption{\small \label{fig:randomised}
		unobservability of deviation in mixed strategy with discount factor $\lambda$}
\end{figure*}

For pure strategies, a deviation by a player can be observed immediately: s/he simply plays a different action than the action defined by the strategy profile assigned by the leader.
Let us now consider a reward and punish strategy for the simple game depicted in Figure \ref{fig:randomised}.
In this example, player~$1$ owns vertices~$1$,~$2$,~$3$ and ~$4$. Leader owns vertices denoted by $l_1$ and $l_2$. 
Rewards are given on the edges and they are in the order (leader, player~$1$). 
When extending the concepts from the previous sections to mixed strategies, the optimal leader strategy profile would be to ask the player~$1$ to play to vertex $l_1$ with a $10\%$ chance, and to $l_2$ with a $90\%$ chance. When player~$1$ follows his strategy, the leader pledges to take an edge from $l_1$ to the vertex~$2$.
While, if player~$1$ deviates at vertex~$1$, leader would harm him by taking an edge from $l_1$ to the vertex~$3$.

The expected reward for the leader would be $8.9 \lambda$, while the expected reward of player~$1$ would be~$0$. Player~$1$ does not benefit from deviation, as, upon deviation, the leader would start to harm him.
In particular, she plays to the vertex~$3$ from $l_1$.

The catch in this concept is that, with normal observational power (where the players can only observe vertices and actions), the leader (and other players in a multi-player game) would only be able to observe \emph{which} action has been taken, but not \emph{why}.
The leader (and other players) cannot distinguish whether the player~$1$ has moved to $l_1$ because he conducted a fair experiment with a $10\%$ chance to move to $l_1$, whose outcome was to move there, or because he simply moved there (with a $100\%$ chance) under deviation from the assigned strategy to improve his payoff.

To be able to distinguish compliance from deviation in mixed strategies, we would therefore need a stronger observation model, where the randomised decision (in our example, the decision to play to $l_1$ with a $10\%$ chance) or the random experiment itself can be observed. Under such an extended observation model, deviation can be observed and we briefly discuss why the results from the previous sections extend to mixed strategies when we assume this observational power.

Also, these temporal dependencies are \emph{not} common in the definition of Nash equilibria. This is also unsurprising when given their origin in the normal-form games\cite{Nash01011950}, where only a single move is played and the concept of history and temporal order of cause and effect does not apply. For us, the concept of observability of deviation by a player outweighs the generality of randomised strategies. 

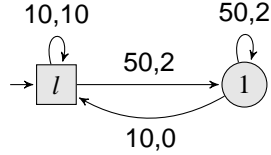
\begin{figure*}
	\centering
	\begin{tikzpicture}[->,>=stealth',shorten >=1pt]
	\tikzstyle{vertex}=[circle,fill=black!10,minimum size=17pt,inner sep=0pt,font=\sffamily\small\bfseries]
	\tikzstyle{vertex1}=[rectangle,fill=black!10,minimum size=15pt,inner sep=0pt,font=\sffamily\small\bfseries]
	\node (4) at (0.25,1) {};
	\node (1) at (1,1) [vertex1,draw]{$l$} ; 	
	\node (2) at (3.5,1) [vertex,draw] {$1$};
	
	\path[every node/.style={font=\sffamily\small}]
	(4) edge [right] node[] {} (1)
	(1) edge [right] node [above] {50,2} (2) 
	edge [loop above] node[] {10,10} (1)      
	(2) edge [bend left] node [below] {10,0} (1)
	edge [loop above] node[] {50,2} (2); 
	
	\end{tikzpicture}\vspace*{-2mm}
	\caption{\small \label{fig:mixedstrategy1}
		leader benefits from memory in mixed strategies}
\end{figure*}

The above argument driven by the unobservability of deviation in mixed strategies made us focus only on the pure strategies.
However, an alternative to this restriction is to lift the restriction of our observational power:
instead of observing the outcome of a decision, we observe the decision itself. Note that this would imply an uncountable set of possible actions, as encoded in the different selected probability distribution over actions, which are possible in every vertex. To justify making this observable, one might think of externalising how to resolve the probabilities, say, by a highly trusted third party. Also, note that allowing for mixed strategies does not remove the usefulness of memory.
In the example from Figure \ref{fig:mixedstrategy1} (player~$1$ owns vertex~$1$, leader owns vertex~$l$ and rewards are in the order leader, player~$1$), when in vertex 1, the leader can only assign an equilibrium strategy to player~$1$, which is not worse for player 1 than staying in vertex 1. 
Initially (that is, on the empty history), however, she does not have to take the interest of player 1 into account and can progress to vertex 1 with probability~$1$.
With this motivation in mind, we define \emph{mixed strategies}, which are functions $\sigma_p: (VA)^*V_p \rightarrow \dist(A)$ from initial sequences of plays that end in some vertex of player $p$ to a distribution over the actions in $A$.
This implies re-writing the expected reward for player $p$ as follows. We use
$\mathbb E_{p}(\sigma,v) = $ $\sum\limits_{a \in A} \sigma(v)(a) \cdot \Big(t_p(v,a)  + \lambda \mathbb E_{p}(\sigma,T(v,a))\Big)$ to denote the payoff for player $p$ when starting at vertex $v$. We again have $\mathbb E_{p}(\sigma)=\sum_{v \in V} \Delta(v)\mathbb E_{p}(\sigma,v)$.

Corollary \ref{cor:optimalPunishment} establishes that it suffices to focus only on the reward and punish strategy profiles.
This implies a simple constraint system for extended memory games: no player (except for the leader in leader equilibria) may reach a position, where a player would benefit from changing her strategy in a reward and punish strategy profile (that is assigned by the leader). Thus, at every vertex $v$ of the extended memory game (with memory $m$), it must hold that
$\mathbb E_{p}(v,m) \geq  \mathbb E_{p,2}(v).$ 
Here, $\mathbb E_{p}(v,m)$ is the expected reward for player $p$ at vertex $v$ in extended memory game and $\mathbb E_{p,2}(v)$ is the expected reward for player $p$ at vertex $v$ in two-player game that would result if player $p$ chooses to deviate at vertex $v$.

We can again use a non-deterministic approach to solve the related decision problem.
We can start by guessing a probability distribution at each vertex $v_{i}$ on all its outgoing actions, and guess, for each action, a target memory value. Once these distributions are fixed, we can again solve the resulting linear equation system, and simply check that it satisfies the constraints from above and meets the required threshold value. Unlike the pure case, where the existence of an optimal solution is implied by the existence of a finite set of possible strategies, we have to provide an argument for the existence of an optimal strategy profile with given memory bound in this setting. According to the constraint system from above, the leader assigns probabilities to the actions and selects the memory updates.
If the resulting system complies with the first set of constraints, then it is a Nash (resp.\ leader) equilibrium. 
Technically, the converse (only if) does not hold, as these constraints only need to be satisfied by the reachable vertices. We could, however, require the same for unreachable vertices without excluding relevant solutions.

\begin{theorem}
	\label{theo:polyhedra}
	For multi-player DSGs with perfect observation and predefined memory an optimal leader strategy profile exists.	
\end{theorem}

\begin{pproof}
First, we know that a strategy profile that satisfies the constraints exists (c.f., Section \ref{sec:strange}).
Further, to see that an \emph{optimal} strategy profile exists, we look at the reward obtained at the different probabilistic transitions. That is, we consider the reward obtained on the different probabilities assigned on different transitions. 
We define the payoff vector as a direct function on the probability assigned on the transitions and the strategy profile as the set 
$\mathcal D$ (for decisions) of probability vectors over actions, or a finite dimensional closed subset of 
$[0,1]^n$ for some $n \in \mathbb N$. This set of probability distributions over the possible actions gives the expected payoff for all players at all positions of the extended memory game graph (game graph with memory of pre-defined size $m$) and is defined by the memory copies at all vertices. 
The resultant payoff for all players at all vertices of the extended game graph is, thus, again a subset of a finite dimensional product of closed and bounded intervals, referred to as $\mathcal P$ (for payoff). The intervals are bounded because, if $p$ defines the maximal absolute value of any of the individual payoffs in the discounted sum game, then every payoff must be in the interval $[-\frac{p}{1-\lambda},\frac{p}{1-\lambda}]$. Given a strategy profile, represented by a $\overrightarrow{d} \in \mathcal D$, we can compute the payoffs, represented by a vector $\overrightarrow{p} \in \mathcal P$. We represent this by a valuation function 
$\mathsf{val}: \mathcal D \rightarrow \mathcal P$, that maps each probability vector to a payoff vector.
The valuation function is continuous: if the decision vector $\mathcal D$ changes only marginally, then the payoff vector $\mathcal P$ changes only marginally, too. Thus, if we fix an $\varepsilon > 0$ then we can first choose a natural number $l$, such that $\sum_{i=l}^\infty \lambda^i p < \varepsilon$, and then choose a $\delta \in ]0,1[$ such that the change between two consecutive probabilities that is given by 
$l\big(1-(1-\delta)^l\big) <  \frac{\varepsilon}{pl}$ is only marginal.
Then, if the absolute sum of changes of all probabilities is below $\delta$, we can estimate the difference by
$\sum_{i=0}^\infty 2\lambda^i p \big(1-(1-\delta)^i\big)$.
For the estimation of this difference, assume that we start with the probability vector $\overrightarrow{d}_m$, which is the point-wise minimum of $\overrightarrow{d}$ and $\overrightarrow{d}'$.
Then the difference can be estimated by choosing the joint actions with the probability described in $\overrightarrow{d}_m$, and simply marking the positions with the missing probability (the difference between the sum of the probabilities reflected in $\overrightarrow{d}_m$ and $1$ at every position in the extended game) as deviation. This difference is bounded by $\delta$.

The likelihood of being in a state where \emph{no} difference has occurred so far is, after $i$ rounds, $\geq (1-\delta)^i$.
The likelihood that a difference has occurred so far can therefore be estimated by $\big( 1 - (1-\delta)^i\big)$.
Using this estimation, we can estimate the difference,\\[1ex]
$\sum_{i=0}^{l-1} 2\lambda^i p\big(1-(1-\delta)^i\big)
+ \sum_{i=l}^\infty 2\lambda^i p \big(1-(1-\delta)^i\big)$
$<
\sum_{i=0}^{l-1} 2 p\big(1-(1-\delta)^l\big)
+ 2 \varepsilon < 4 \varepsilon$,\\[1ex]
where the first inequality uses the definition of $l$, $\lambda^i \leq 1$, and $1-(1-\delta)^i < 1- (1-\delta)^l$, while the second estimation uses the definition of $\delta$.
Thus, $\forall \varepsilon>0$ $\exists\delta>0$ such that $\| \overrightarrow{d} - \overrightarrow{d}'\| < \delta$ implies $\| \mathsf{val}(\overrightarrow{d}) - \mathsf{val}(\overrightarrow{d}') \| < 4\varepsilon$.
The subset $\mathcal C \subseteq \mathcal P$ of the set of payoffs that comply with the constraint system is obviously still closed, as it is still a product of finitely many closed intervals. (Only the lower bound of these intervals may have changed.)
As $\mathsf{val}$ is continuous, the preimage $\mathcal D'$ of the closed and bounded set $\mathcal C$ is closed and bounded.
When $\mathsf{val}$ is restricted to $\mathcal D'$, then the maximum w.r.t.\ the value of the leader in the initial state exists. That is, the supremum is taken for some value.
\end{pproof}

\section{Conclusions}
We have established the usefulness of memory in obtaining optimal leader strategy profiles in discounted sum games. Strategy profiles could be formed from memoryless, bounded memory or infinite memory strategies. Unsurprisingly, more memory can help. Our simple example from Figure \ref{fig:finitegain} had shown that there is no upper bound that could be inferred from the structure of the game on the memory needed for an optimal strategy profile. We observed that in some cases even infinite memory is needed (c.f., Theorem \ref{theo:memorybound}, Theorem \ref{theo:memorybound2} and Theorem \ref{theo:memorybound-NE}). We have argued that (and why) the detectability of deviation made the restriction to pure strategies a natural choice. We showed that the related decision problem (Is there a Nash resp.\ leader equilibrium that provides a payoff that meets or exceeds a given threshold?) is NP-complete. We have also discussed the extension to mixed strategies with bounded memory and the extension of the observation model that is needed to make such strategies reasonable.
Possible future work could be to implement our nondeterministic approach for solving these games in SMT solvers like Yices \cite{DBLP:conf/cav/MouraDS07,yices} and see how well they perform on small examples.

\nocite{*}
\bibliographystyle{eptcs.bst}

\providecommand{\urlalt}[2]{\href{#1}{#2}}
\providecommand{\doi}[1]{doi:\urlalt{http://dx.doi.org/#1}{#1}}

\end{document}